\documentclass[12pt]{article}
\usepackage{amsfonts}
\usepackage{times}
\def\End{\mathrm{End}}\def\C{\mathbb{C}}
\newcommand{\beq}{\begin{equation}}\newcommand{\eeq}{\end{equation}}
\newcommand{\ba}{\begin{array}}\newcommand{\ea}{\end{array}}
\newcommand{\beqa}{\begin{eqnarray}}\newcommand{\eeqa}{\end{eqnarray}}

\newcounter{Theorem}\newcounter{Remark}\newcounter{Definition}
\newcounter{Example}\newcounter{Exercise}

\newenvironment{Theorem}[1][\bf Theorem \arabic{Theorem}]{%

        \refstepcounter{Theorem}\noindent\textbf{#1.}${}$\hspace{1pt}${}$\it}{}

\newenvironment{Lemma}[1][\bf Lemma \arabic{Theorem}]{%
        
        \refstepcounter{Theorem}\noindent\textbf{#1.}${}$\hspace{1pt}${}$\it}{}
\newenvironment{Corollary}[1][\bf Corollary \arabic{Theorem}]{%
        
        \refstepcounter{Theorem}\noindent\textbf{#1.}${}$\hspace{1pt}${}$\it}{}
\newenvironment{Remark}[1][\bf Remark \arabic{Remark}]{%
        
        \refstepcounter{Remark}\noindent\textbf{#1.}${}$\hspace{1pt}${}$}{}

\newenvironment{Proof}[1][\bf Proof.]{\noindent\textrm{#1}${}$\hspace{7pt}${}$}{\nolinebreak${}$\hfill $\Box$}

\begin{document}
\title{ Cremmer-Gervais Quantum Lie Algebra}
\author{Oleg Ogievetsky\footnote{On leave of absence from P.N. Lebedev Physical Institute, 
Theoretical Department, Leninsky prospekt 53, 119991 Moscow, Russia}\\[.6em] 
{\it Centre de Physique Th\'eorique\footnote{Unit\'e Mixte de Recherche (UMR 6207) du CNRS et des 
Universit\'es Aix--Marseille I, Aix--Marseille II et du Sud Toulon -- Var; laboratoire affili\'e 
\`a la FRUMAM (FR 2291)}, Luminy, 13288 Marseille, France } \\[1.8em] 
Todor  Popov \\[.6em] 
\it Institute for Nuclear Research and Nuclear Energy, \\ 
\it Bulgarian Academy of Sciences, Sofia, 
BG-1784, Bulgaria}
\date{}
\maketitle
\begin{abstract} \noindent We describe a quantum Lie algebra based on the Cremmer-Gervais R-mat\-rix. The algebra arises upon a restriction of an
infinite-dimensional quantum Lie algebra.
\end{abstract}

\bigskip

\section{Introduction}

The notion of a quantum Lie algebra is a modification of the notion of a Lie algebra. Quantum Lie algebras arise as the algebras generated by the quantum analogs of
vector fields in the framework of the bicovariant differential calculus on  quantum groups \cite{W} (for an introduction see e.g. \cite{AC}).
Many constructions from the theory of Lie algebras can be generalized for quantum Lie algebras (for example, the standard complex, BRST operator {\em etc.} \cite{IO, IO2,GIO}).

\vskip .2cm
In this Note we outline the quantum Lie algebra having the so called Cremmer-Gervais R-matrix \cite{CG} as the braid matrix. This can be
seen as a first step in constructing the BRST operator for the bicovariant differential calculus based on the Cremmer-Gervais R-matrix.

 \section{Quantum Lie Algebra}

The bicovariant differential calculus is characterized by functionals $\chi_i$ and $ f^i_j$ on a Hopf algebra $\cal A$ ( ``the algebra
of functions on a quantum group'') satisfying the relations
\beq\label{bcc}\begin{array}{lll}
&\chi_i \chi_j -  \sigma^{kl}_{ij} \chi_k \chi_l =  C^k_{ij} \chi_k\ ,
&\sigma^{kl}_{ij} f^a_k f^b_l = f^k_i f^l_j\sigma^{ab}_{kl} \ ,\\[1em]
 &\sigma^{kl}_{ij} \chi_k f^a_l + C^l_{ij} f^a_l =f^k_i f^l_j C^a_{kl}+f^a_i \chi_j\ ,
&\chi_i f^a_j = \sigma^{kl}_{ij} f^a_k \chi_l \  .\end{array}\eeq
Here the structure constants  $C^i_{jk}$ and the braid matrix $\sigma^{ij}_{kl}$ ($C^i_{jk}$ and $\sigma^{ij}_{kl}$ are subject to certain conditions, see below) are
such that $C^i_{jk}= \chi_k (M^i_j)$ and $\sigma^{ij}_{kl}=f^i_l (M^j_k)$,
where the matrix $M\in \cal A$ is given by the right coaction on the space of left-invariant forms
$$\Delta_R(\omega^i)= \omega^j \otimes M^i_j\ , \qquad \qquad  M^i_j \in {\cal A }\ , \qquad \omega^i \in \Gamma \ .$$
The algebra (\ref{bcc}) endowed with the comultiplication $\Delta$, counit $\epsilon$ and antipode $S$,
\beq\label{hopf}\ba{lclcl}
\Delta f^i_j = f^i_k \otimes f^k_j \ ,&& \epsilon(f^i_j)= \delta^i_j\ , && S (f^i_k) f^k_j = \delta^i_j=  f^i_k S(f^k_j)\ ,  \\[.6em]
\Delta \chi_i = 1 \otimes \chi_i + \chi_j \otimes f^j_i\ , && \epsilon(\chi_i)=0\ ,&& S(\chi_i)= - \chi_j S(f^j_i)\ , \ea\eeq
becomes a Hopf algebra which we will be  denote by ${\cal L}$.  The subalgebra generated by $\chi_i$ is called {\em quantum Lie algebra}.

The relations for $\cal L$ can be written in a concise way with the help of a single R-matrix \cite{B}. Let us make a convention that
the small  indices $i, j, \ldots ,k$ run over a set ${\cal I }$ and the capital indices $I,J, \ldots, K $ run over the set ${\cal
I}_0:=0\cup {\cal I} $. Denote by $\hat{R}$ and $T$ the following matrices 
\beq\label{conv}\hat{R}^{I j}_{k L}=\left(\ba{cc}\delta^j_k& C^j_{kl}\\ 0& \sigma^{ij}_{kl}\ea \right)\
, \qquad \hat{R}^{I 0}_{0 L} = \delta^I_L\ , \qquad \qquad T^I_J=
\left(\ba{cc}1 & \chi_j \\ 0& f^i_j\ea \right)\ ,\eeq i.e.,
$\hat{R}^{ij}_{kl}= \sigma^{ij}_{kl} \, , \hat{R}^{0j}_{kl} =
C^j_{kl}\, , \hat{R}^{0A}_{B0} = \delta^A_B \, ,  \hat{R}^{A0}_{0B}
= \delta^A_B$ and  $ T^i_j = f^i_j \, , T^0_j=\chi_j \, , T^0_0=1$
and all others entries are equal to zero. Suppose now that $R$ is a solution of the Yang-Baxter equation
$$\hat{R}_{12}\hat{R}_{23}\hat{R}_{12}=\hat{R}_{23}\hat{R}_{12}\hat{R}_{23} \ .$$
Then the Hopf algebra relations  (\ref{bcc}) and  (\ref{hopf}) are equivalent to
\beqa
\hat{R}_{IJ}^{KL}T_K^A T_L^B=T_I^K T_J^L \hat{R}^{AB}_{KL}\ , \quad \ && \Delta T^I_J = T^I_ K \otimes T^K_J\ , \nonumber  \\[1em] 
\ S(T^I_K)T^K_J = \delta^I_J= T^I_K S(T^K_J)\ , && \epsilon(T^I_J)=\delta^I_J\nonumber \ .\eeqa

The Yang-Baxter relation for $\hat{R}$ implies, for the components $\sigma^{ij}_{kl}$ and $C^j_{kl}$,
\beq\!\ba{ll}
C^s_{ni}  C^m_{sj} -  \sigma^{kl}_{ij} C_{nk}^s C_{sl}^m =  C^k_{ij}  C^m_{nk}\ ,& \sigma^{kl}_{ij} \sigma^{as}_{nk}\sigma^{bm}_{sl}=\sigma^{ks}_{ni}\sigma^{lm}_{sj}\sigma^{ab}_{kl}\ ,\\[1em]
\sigma^{kl}_{ij} C^s_{nk} \sigma^{am}_{sl} + C^l_{ij} \sigma^{am}_{n l}
=\sigma^{ks}_{ni}\sigma^{lm}_{sj}  C^a_{kl} + \sigma^{as}_{ni} C^m_{sj}\ , &
C^s_{ni} \sigma^{am}_{sj} = \sigma^{kl}_{ij} \sigma^{as}_{nk} C^m_{sl} \ .\ea \eeq
Here the first relation  is the ``braided'' Jacobi identity and the second one is simply the braid relation for $\sigma$,
 $\sigma_{23}\sigma_{12}\sigma_{23}=\sigma_{12}\sigma_{23}\sigma_{12}$.

\vskip .2cm
Given a braid matrix $\sigma$ it is natural to ask if non-zero structure constants $C^k_{ij}$ consistent with $\sigma$ exist or is there
a non-trivial quantum Lie algebra structure compatible with $\sigma$.
As we have seen this question is equivalent to finding a suitable extension (\ref{conv}) of the R-matrix $\sigma$.

In this Note we obtain an infinite-dimensional R-matrix which upon restrictions yields finite-dimensional
quantum Lie algebras compatible with the Crem\-mer--Gervais $R$-matrix \cite{CG}.

\section{Cremmer-Gervais extended}

We apply the elegant method used in \cite{SU} and then in \cite{EH} where the Yang-Baxter operators are realized as operators in a certain space of functions.
{}Finite-dimensional R-matrices arise upon a restriction of the operator domain  to an appropriate invariant finite-dimensional subspace, such as the space
of polynomials of bounded degree.

\vskip .2cm
{}For a ring $K$, let $K (x)$ be the ring of rational functions in $x$ with coefficients in $K$. An  endomorphism of $K$ extends to an endomorphism of
$K (x)$ (which acts only on the coefficients of a rational function). Having an endomorphism $\phi \in \End \, \mathbb{C} (x,y)$, introduce
$\phi_{12} \in \End \, \mathbb{C} (x,y,z)$ considering $ \mathbb{C}(x,y,z)$ as $\mathbb{C}(x,y)(z)$. In the same vein, $\phi_{13}\in \End \, \C(x,z)(y)$
and $\phi_{23}\in \End \, \C(y,z)(x)$ and the functional Yang-Baxter equation reads $\phi_{12}\phi_{13}\phi_{23}=\phi_{23}\phi_{13}\phi_{12}$ .

\vskip .3cm
Given a rational function $F(x,y)$ with  series expansion (around 0) $F(x,y)= \sum_{i,j \in \mathbb Z} F_{i,j} x^i y^j$,
define the operation $reg_{x,y}$ which maps $F(x,y)$ to the non-sin\-gu\-lar part $f(x,y)$ of its expansion,
\[ f(x,y)=reg_{x,y}F(x,y):= \sum_{i,j \geq 0} F_{i,j} x^i y^j \ .\]

\vskip .3cm
\begin{Theorem}$\ $  Let $\hat{R}$ be the following linear operator in $ \End \, \C(x,y)$
\[ \hat{R} =  P  + \beta \frac{y}{x-y}(P-I) reg_{x,y} + \frac{C}{x} eval_{x=0} (P-I)reg_{x,y}\ ,\]
$\beta=1- q^{-2}$ and $C$ are arbitrary constants. Here $I$ stands for the identity operator, $P$ for the permutation $(P F)(x,y) =F(y,x)$ and $eval_{x=0}$ is
the evaluation at $x=0$; in other words, for an arbitrary $F(x,y)\in \C(x,y)$ the result of the action of the operator $\hat{R}$ reads
\beq\label{Rh}(\hat{R}F)(x,y)= F(y,x) + \beta y\frac{f(y,x)-f(x,y)}{x-y} + C \frac{f(y,0)-f(0,y)}{x} \ .  \eeq
The operator $\hat{R}$ satisfies the braid equation $\hat{R}_{12}\hat{R}_{23}\hat{R}_{12}=\hat{R}_{23}\hat{R}_{12}\hat{R}_{23}$.
\end{Theorem}

\vskip .3cm
\begin{Proof} The braid equation for the operator $\hat{R}$ is equivalent to the Yang-Baxter equation  $R_{12}R_{13}R_{23}= R_{23}R_{13}R_{12}$ for $R=P\hat{R}$.
The operator $r$ defined through $R=I+r$ reads
\beq \label{rcl}r= \beta \frac{x}{x-y}(P-I)reg_{x,y} + \frac{C}{y} eval_{y=0} (I-P) reg_{x,y} \ .\eeq
It acts on a function $F(x,y)$ as follows: $(rF)(x,y)= \beta (\rho F)(x,y)+C  ({\mathfrak{s}} F)(x,y)$, where
\beq  (\rho F)(x,y)= x \frac{f(y,x)-f(x,y)}{x-y}\ ,\ \  ({\mathfrak{s}} F)(x,y)=\frac{f(x,0)-f(0,x)}{y}\ .\eeq
Note that $r$ sends a polynomial $f(x,y)$ to the sum of a polynomial in $x$ and $y$ (the term with the coefficient $\beta$ ) and a polynomial in
$x$ (with the coefficient $\frac{C}{y}$).

\vskip .2cm
The first step is to check that $r$ is a classical r-matrix.

\vskip .3cm
\begin{Lemma}$\ $ The operator $r$ satisfies the classical Yang-Baxter equation
\beq [r_{12},r_{13}]+ [r_{12},r_{23}]+ [r_{13},r_{23}]=0\ . \label{clyab}\eeq
\end{Lemma}

\vskip .3cm
\noindent
{\bf Proof of the lemma.} The operator in the left hand side of (\ref{clyab}) depends only on the regular part of a  function $F(x,y,z)\in \C(x,y,z)$ therefore
it is enough to check the assertion on an arbitrary polynomial 
 $f(x,y,z)\in \C[x,y,z]$. 
Since the coefficients $\beta$ and $C$ are arbitrary,
the classical Yang-Baxter equation for $r$ splits into three components. The component proportional to $\beta^2$ is the classical Yang-Baxter
equation for $\rho$; it is satisfied: $\rho$ is the classical Cremmer-Gervais $r$-matrix \cite{EH,OP}. Next, a straightforward verification shows that
$$\begin{array}{c}\rho_{13}{\mathfrak{s}}_{23}=0\ ,\ \rho_{23}{\mathfrak{s}}_{13}=0\ ,\ \rho_{23}{\mathfrak{s}}_{12}=0\ ,\ [{\mathfrak{s}}_{12},\rho_{13}]
+{\mathfrak{s}}_{12}\rho_{23}=0\ ,\\[1em] [\rho_{12},{\mathfrak{s}}_{13}]+ {\mathfrak{s}}_{13}\rho_{23}- {\mathfrak{s}}_{23}\rho_{13}
+ [\rho_{12},{\mathfrak{s}}_{23}]=0\end{array}$$
on polynomials. The sum (with corresponding signs) of these equalities is the component proportional to $\beta C$. Finally, a straightforward
verification shows that
$${\mathfrak{s}}_{23}{\mathfrak{s}}_{12}=0\ ,\ {\mathfrak{s}}_{23}{\mathfrak{s}}_{13}=0\ ,\ {\mathfrak{s}}_{13}{\mathfrak{s}}_{23}=0\ ,\ [{\mathfrak{s}}_{12}
,{\mathfrak{s}}_{13}]+{\mathfrak{s}}_{12}{\mathfrak{s}}_{23}=0$$
on polynomials and the classical Yang-Baxter equation for the operator ${\mathfrak{s}}$ (the component, proportional to $C^2$) follows. \hfill$\Box$

\vskip .3cm
The  Yang-Baxter equation for $R=I+r$ holds true if  the operator $r$ satisfies the classical Yang-Baxter equation and the Yang-Baxter equation
 \beq r_{12}r_{13}r_{23}=r_{23}r_{13}r_{12}.   \label{ybfr}\eeq
To check  this identity on an arbitrary function $F(x,y,z)\in
\C(x,y,z)$ it is again enough to check it on an arbitrary polynomial
function $f(x,y,z)\in\C[x,y,z]$. Now (\ref{ybfr}) splits into four
components.  The component proportional to $\beta^3$ vanishes
($\rho$ satisfies a stronger equation, see \cite{OP}). Next, a
direct verification shows that
\beqa & {\mathfrak{s}}_{12} \rho_{13} \rho_{23}=0\ ,\quad 
\rho_{12} \rho_{13}{\mathfrak{s}}_{23}=0\ ,\quad 
\rho_{23}{\mathfrak{s}}_{13} \rho_{12}=0\ ,  & \nonumber
\\[.6em]
&\rho_{23} \rho_{13}{\mathfrak{s}}_{12}=0\ ,\quad \rho_{12}{\mathfrak{s}}_{13} \rho_{23}={\mathfrak{s}}_{23} \rho_{13} \rho_{12} &  \nonumber
\eeqa
on polynomials; the vanishing of the component proportional to $\beta^2 C$ follows. Finally, each term in the components, proportional to $\beta C^2$,
\beqa{\mathfrak{s}}_{12}{\mathfrak{s}}_{13}\rho_{23}=0\ , &\ {\mathfrak{s}}_{12}\rho_{13}{\mathfrak{s}}_{23}=0\ , & \ \rho_{12}{\mathfrak{s}}_{13}{\mathfrak{s}}_{23}=0\ ,\nonumber 
\\[.6em] \rho_{23}{\mathfrak{s}}_{13}{\mathfrak{s}}_{12}=0\ , & \ {\mathfrak{s}}_{23}\rho_{13}{\mathfrak{s}}_{12}=0\ , & \ {\mathfrak{s}}_{23}{\mathfrak{s}}_{13}\rho_{12}=0
\nonumber\ ,\eeqa
and $C^3$,
$${\mathfrak{s}}_{12}{\mathfrak{s}}_{13}{\mathfrak{s}}_{23}=0\ ,\ {\mathfrak{s}}_{23}{\mathfrak{s}}_{13}{\mathfrak{s}}_{12}=0\ ,$$
vanishes separately, which ends the proof of the theorem.\end{Proof}

\vskip .3cm
Let $V=\bigoplus_{i=0}^{n}\C e_i$ be a finite-dimensional vector space of functions $\frac{p(x)}{x}$ where $p(x)$ is a polynomial of degree not higher than $n$.
Identify $V\otimes V$ with the space of functions $\frac{p(x,y)}{x y}$ where $p(x,y)$ is a polynomial of degree not higher than $n$ in $x$ and not higher than $n$
in $y$.  The space $V\otimes V$ is stable under the action of the operator $\hat{R}$. The  matrix  of the restricted operator
$\hat{R}(e_K\otimes e_L) = \sum_{I,J=0}^n e_I \otimes e_J \hat{R}^{IJ}_{KL}$ (which we denote again by $\hat{R}$) is given by
 \[  \hat{R}(x^{K-1} y^{L-1})= \sum_{I,J=0}^n  \hat{R}^{IJ}_{KL} x^{I-1} y^{J-1}\ ;   \qquad I,J,K,L=0, \ldots, n\ .\]
The non-vanishing entries of the matrix $\hat{R}^{IJ}_{KL}$ read as follows
\beq\hat{R}^{0J}_{K0}=\hat{R}^{J0}_{0K} =  \delta^J_K\ ,
\qquad \hat{R}^{0j}_{kl} = C^j_{kl}=C  ( \delta^1_l \delta^j_k -\delta^1_k \delta^j_l )\ , \eeq
\beq\label{eCG}\hat{R}^{ij}_{kl} =  \delta^{i}_{l} \delta^{j}_{k}+(1-q^{-2})\left( \sum_{k \leq s < l}-\sum_{l \leq s < k} \right)  \delta^{i}_{s} \delta^{j}_{k+l-s}\ ,\eeq
$ i,j,k,l=1, \ldots, n$.

\vskip .1cm
The latter submatrix $\hat{R}^{ij}_{kl}$ is the member (with $p=1$) of the   Cremmer-Gervais family of non-unitary R-matrices
\beq\label{CG} (\hat{R}_{CG,p})^{ij}_{kl} =  p^{k-l} \delta^{i}_{l} \delta^{j}_{k}+(1-q^{-2}) \left( \sum_{k \leq s < l}
-  \sum_{l \leq s < k}   \right) p^{k-s} \delta^{i}_{s} \delta^{j}_{k+l-s} \ .\eeq

\vskip .2cm
We sum up these results in the following corollary.

\vskip .3cm
\begin{Corollary}$\ $ The above finite-dimensional restriction of the operator $\hat{R}$, defined by (\ref{Rh}), gives rise to a quantum Lie algebra associated with the $p=1$
member of the family of non-unitary Cremmer-Gervais R-matrices. The non-zero structure constants $C^{j}_{kl}$ are all equal to $\pm C$ (the constant $C$ can be set
to 1 by rescalings),
\[  C^j_{j1}=-C^{j}_{1j}  = C\ ,  \qquad \qquad  j= 1 \ldots n \ .\]
\end{Corollary}

\vskip .2cm
\begin{Remark}$\ $ Our treatment is  an extension of the construction of \cite{EH}  in which the finite-dimensional Cremmer-Gervais matrices arise upon restrictions of
infinite-dimensional functional R-matrices to the spaces of polynomials. The  boundary (unitary) Cremmer-Gervais solution of the Yang-Baxter equation can be treated
along the same lines \cite{EH}. The boundary Cremmer-Gervais R-matrix as well gives rise to a quantum Lie algebra which will be described elsewhere.
\end{Remark}

\section*{Acknowledgements}
 The work was partially supported by the ANR project GIMP No.ANR-05-BLAN-0029-01. T.P. thanks
Centre de Physique Th\'eorique, Luminy for the hospitality. The work of T.P. was also partially supported
by the European Operational program HRD through contract BGO051PO001/07/3.3-02/53 with the
Bulgarian Ministry of Education.


\begin{thebibliography}{10}
\bibitem{W} S. L. Woronowicz, Comm. Math. Phys. {\bf 122} (1989) 125-170.
\bibitem{AC} P. Aschieri and L. Castellani, Int. J. Mod. Phys. A {\bf 8} (1993) 1667.
\bibitem{IO} A. P. Isaev and O. V. Ogievetsky, Theoretical and Math. Physics {\bf 129} (2001) 1558-1572.
\bibitem{IO2} A. P. Isaev and O. V. Ogievetsky, Int. J. Math. Phys. A {\bf 19} (2004) 240-247.
\bibitem{GIO} V. G. Gorbounov,  A. P. Isaev and O. V. Ogievetsky,
Theoretical and Math. Physics {\bf 139} (2004) 473-485.
\bibitem{B} D. Bernard, Phys. Lett. {\bf B260} (1991) 389-393.
\bibitem{CG} E. Cremmer and J.-L. Gervais, Comm. Math. Phys. {\bf 134 } (1990) 619-632.
\bibitem{SU} Y. Shibukawa and K. Ueno, 
Lett. Math. Phys. {\bf 25} (1992) 239-248.
\bibitem{EH} R. Endelman and T. Hodges,  Lett. Math. Phys. {\bf 52} (2000) 225--237.
\bibitem{OP} O. Ogievetsky and T. Popov, {\em R-matrices in Rime}; ArXiv: 0704.1947 [math.QA]
\end{thebibliography}
\end{document}